\documentclass[preprint]{aastex}

\let\url\relax 
\usepackage{hyperref}

\usepackage{lineno}
\usepackage{amsmath}
\usepackage{lscape}
\usepackage{color}

\newcommand{\fermi}{\emph{Fermi}}
\newcommand{\DG}{$^{\circ}$}
\newcommand{\mypsr}{PSR J1227$-$4853}
\newcommand{\Per}{$P$}
\newcommand{\Pd}{$\dot{P}$}
\newcommand{\Ed}{$\dot{E}$}

\begin{document}

\title{Discovery of Gamma-ray Pulsations from the Transitional Redback PSR J1227$-$4853}

\author{
T.~J.~Johnson\altaffilmark{1,2}, 
P.~S.~Ray\altaffilmark{3,4}, 
J.~Roy\altaffilmark{5,6,7}, 
C.~C.~Cheung\altaffilmark{3}, 
A.~K.~Harding\altaffilmark{8}, 
H.~J.~Pletsch\altaffilmark{9,10}, 
S.~Fort\altaffilmark{9,10}, 
F.~Camilo\altaffilmark{11},
J.~Deneva\altaffilmark{12},
B.~Bhattacharyya\altaffilmark{5},   
B.~W.~Stappers\altaffilmark{5},
M.~Kerr\altaffilmark{13}
}
\altaffiltext{1}{College of Science, George Mason University, Fairfax, VA 22030, resident at Naval Research Laboratory, Washington, DC 20375, USA}
\altaffiltext{2}{email: tyrel.j.johnson@gmail.com}
\altaffiltext{3}{Space Science Division, Naval Research Laboratory, Washington, DC 20375-5352, USA}
\altaffiltext{4}{email: Paul.Ray@nrl.navy.mil}
\altaffiltext{5}{Jodrell Bank Centre for Astrophysics, School of Physics and Astronomy, The University of Manchester, Manchester M13 9PL, UK}
\altaffiltext{6}{National Centre for Radio Astrophysics, Tata Institute of Fundamental Research, Pune University Campus, Pune 411 007, India}
\altaffiltext{7}{email: jayanta.roy@manchester.ac.uk}
\altaffiltext{8}{NASA Goddard Space Flight Center, Greenbelt, MD 20771, USA}
\altaffiltext{9}{Albert-Einstein-Institut, Max-Planck-Institut f\"ur Gravitationsphysik, D-30167 Hannover, Germany}
\altaffiltext{10}{Leibniz Universit\"at Hannover, D-30167 Hannover, Germany}
\altaffiltext{11}{Columbia Astrophysics Laboratory, Columbia University, New York, NY 10027, USA}
\altaffiltext{12}{National Research Council Research Associate, National Academy of Sciences, Washington, DC 20001, resident at Naval Research Laboratory, Washington, DC 20375, USA}
\altaffiltext{13}{CSIRO Astronomy and Space Science, Australia Telescope National Facility, Epping NSW 1710, Australia}


\begin{abstract}
The 1.69 ms spin period of \mypsr\ was recently discovered in radio observations of the low-mass X-ray binary XSS J12270$-$4859 following the announcement of a possible transition to a rotation-powered millisecond pulsar state, inferred from decreases in optical, X-ray, and gamma-ray flux from the source.  We report the detection of significant (5$\sigma$) gamma-ray pulsations after the transition, at the known spin period, using $\sim$1 year of data from the Large Area Telescope on board the \emph{Fermi Gamma-ray Space Telescope}.  The gamma-ray light curve of \mypsr\ can be fit by one broad peak, which occurs at nearly the same phase as the main peak in the 1.4 GHz radio profile.  The partial alignment of light-curve peaks in different wavebands suggests that at least some of the radio emission may originate at high altitude in the pulsar magnetosphere, in extended regions co-located with the gamma-ray emission site.  We folded the LAT data at the orbital period, both pre- and post-transition, but find no evidence for significant modulation of the gamma-ray flux.  Analysis of the gamma-ray flux over the mission suggests an approximate transition time of 2012 November 30.  Continued study of the pulsed emission and monitoring of \mypsr, and other known redback systems, for subsequent flux changes will increase our knowledge of the pulsar emission mechanism and transitioning systems.
\end{abstract}

\keywords{pulsars: individual (J1227$-$4853)--binary--gamma rays: stars}

\section{INTRODUCTION}\label{intro}
Millisecond pulsars (MSPs) are thought to be old neutron stars that have reached short rotational periods ($\lesssim$ 10 ms) as the result of accretion from a binary companion \citep[e.g.,][]{Alpar82}.  This hypothesis is supported by the detection of X-ray millisecond pulsations from some low-mass X-ray binaries \citep[LMXBs, e.g.,][]{Wij98}.  However, the existence of isolated MSPs such as the first MSP ever discovered, PSR B1937+21 \citep{Backer82}, requires a separate formation channel \citep[e.g.,][]{Ivanova08} or some mechanism by which the binary system can either be disrupted or the companion destroyed.  With the discovery of the first ``black widow'' pulsar \citep[PSR B1957+20,][]{Fruchter88} the theory that the binary companion could be totally or partially ablated by the energetic pulsar wind, once the MSP turned on as a radio pulsar, gained popularity.  However, other black widows detected in the Galactic field before $\sim$2008 \citep[e.g., J2051$-$0827,][]{J2051b,J2051a} were not as energetic and their companions did not appear to be losing mass at a significant rate, complicating the ablation hypothesis.  Several theories have been put forth as to how the ablation proceeds \citep[e.g.,][]{RST89,HG90,LE91,Takata12}, but one problem with this scenario, in general, was the relative lack of black widows outside of globular clusters \citep{King03}.

Black widow systems are characterized by relatively short orbital periods (few hours) and extremely low-mass ($\sim0.02M_{\odot}$) companions while their cousins the redbacks have more massive ($\sim0.2M_{\odot}$) non-degenerate companions and show radio eclipses \citep{Roberts11}.  Since the launch of the \emph{Fermi Gamma-ray Space Telescope} in 2008, radio astronomers have increased the number of known black widow and redback systems outside of globular clusters, and normal MSPs, via targeted observations of unassociated gamma-ray sources with pulsar-like characteristics \citep[e.g.,][]{Ray12}.  There are now at least 18 and 8 black widow and redback systems known, respectively, in the Galactic field compared to 3 and 1 before \fermi.  These discoveries lend credence to the idea that isolated MSPs can be formed via ablation and, with a larger sample outside of the complicated environments of globular clusters, provide an opportunity to better understand this process.

Recently, three redback systems have been observed to transition between rotation-powered MSP and accretion-powered LMXB states, implying that MSPs may undergo multiple transitions into and out of accreting states late in the recycling process and providing the opportunity to study MSP evolution in more detail.  PSR J1824$-$2452I, in the globular cluster M28, underwent a month-long X-ray outburst, with accretion-powered X-ray pulsations, and was no longer detectable as a rotation-powered radio pulsar before dimming and again being seen as a radio MSP \citep{Papitto13}.  \citet{Stappers14} reported the disappearance of radio pulsations from PSR J1023+0038 near the end of June 2013, accompanied by a five-fold increase in gamma-ray flux above 0.1 GeV.  \citet{Paturno13} reported X-ray and optical observations of this MSP and argued that there was evidence for the presence of an accretion disk.  Accretion-powered X-ray pulsations at the spin period were later detected \citep{Archibald14}.  The LMXB XSS J12270$-$4859 has been spatially associated with an unidentified gamma-ray source \citep[2FGL J1227.7$-$4853,][]{deM10,Hill11}.  By analogy with PSR J1023+0038, \citet{Hill11} hypothesized that this might be a transitional system, but without concurrent spectroscopic observations it was unclear if accretion was still occurring.  The situation was clarified in late 2012 when \citet{Bassa13,Bassa14} reported a drastic decrease of the X-ray and optical flux of XSS J12270$-$4859.  They suggested that the source may have transitioned to a rotation-powered pulsar state and noted a possible decrease in the gamma-ray flux of the associated point source, approximately coincident with the estimated transition time.  Radio observations with the Giant Metrewave Radio Telescope discovered a pulsar with a 1.69 ms spin period, in a binary system with an orbital period of 6.9 hours, consistent with the position of XSS J12270$-$4859, giving the source the designation \mypsr\ and firmly establishing that a state change had been observed \citep{J1227Radio,J1227ATEL}.  Once the spin and orbital periods were known, it was possible to detect X-ray pulsations in archival \emph{XMM-Newton} observations before the state transition, but only during sub-luminous accretion disk states \citep{J1227XMM}.

Using $\sim$1 year of data from the \fermi\ Large Area Telescope \citep[LAT,][]{LATpaper} after the transition and the timing solution of \citet{J1227Radio}, we report the detection of pulsed gamma rays from \mypsr\ with a significance of 5.0$\sigma$.  We also characterize the spectral behavior of the source from the start of the mission until 2012 December and revisit searches for variability at the orbital period, which is slightly different than the optical period reported by \citet{Bassa14}.  Finally, we compare our results to previous studies of the gamma-ray emission from this system, discuss the energetics, and provide thoughts on emission models before and after the transition based on our observations.

\section{ANALYSIS AND RESULTS}\label{an}
\subsection{DATA SELECTION AND PREPARATION}\label{prep}
The LAT is a pair-production telescope sensitive to gamma rays with energies from 20 MeV to more than 300 GeV and observes the gamma-ray sky with 2.4 sr field of view.  At 1 GeV, the LAT has a 68\% containment radius of $\sim$1\DG\ and a near on-axis effective area of $\sim$7000 cm$^{2}$.  For a detailed description of the on-orbit performance of the LAT, see \citet{P7}\footnote{See also \url{http://www.slac.stanford.edu/exp/glast/groups/canda/lat\_Performance.htm}.}.  

We selected events from the \texttt{P7REP} LAT data\footnote{\url{http://fermi.gsfc.nasa.gov/ssc/data/analysis/documentation/Pass7REP_usage.html}} corresponding to the \texttt{SOURCE} class recorded between 2008 August 4 and 2014 December 1, with reconstructed directions within 15\DG\ of the position of \mypsr, energies from 0.1 to 100 GeV, and zenith angles $\leq$ 100\DG.  Good time intervals were then selected corresponding to when the instrument was in nominal science operations mode, the rocking angle of the spacecraft did not exceed 52\DG\ or the limb of the Earth did not infringe upon the region of interest, and the data were flagged as good.  Analysis of LAT data was performed using the \fermi\ ScienceTools\footnote{The \fermi\ ScienceTools can be downloaded at \url{http://fermi.gsfc.nasa.gov/ssc/data/analysis/software/}.} v9r34p2.

\subsection{SPECTRAL ANALYSIS}\label{spec}
We first performed a binned maximum likelihood analysis on a $20^{\circ}\times20^{\circ}$ region centered on the source position using the \texttt{P7REP\_SOURCE\_V15} instrument response functions (IRFs) spanning the entire data set.  All sources from the third \fermi\ LAT source catalog\footnote{See \url{http://fermi.gsfc.nasa.gov/ssc/data/access/lat/4yr_catalog/}.} \citep[3FGL,][]{3FGL} within 25\DG\ of \mypsr\ were included in the model of the region and all spectral parameters of sources within 8\DG\ with average significances $\geq\ 10\sigma$ over four years were left free.  We also left free the normalization parameters of sources within 10\DG\ that were flagged as significantly variable, even if they did not pass the average significance cut.  The spectral parameters of all other sources from 3FGL were kept fixed.  We moved the position of the 3FGL source associated with \mypsr\ (3FGL J1227.9$-$4854) to the timing position from \citet{J1227Radio}.  The Galactic diffuse emission was modeled using the \textit{gll\_iem\_v05\_rev1.fits} model while the isotropic diffuse emission and residual background of misclassified cosmic rays were jointly modeled using the \textit{iso\_source\_v05.txt} template\footnote{The diffuse models are available for download at \url{http://fermi.gsfc.nasa.gov/ssc/data/}.}.  The normalization parameters of both diffuse components were left free.

We modeled the spectrum of \mypsr\ as both a single power law (Equation \ref{eq:pl}) and an exponentially-cutoff power law (Equation \ref{eq:ecpl}):
\begin{equation}\label{eq:pl}
\frac{dN}{dE}\ =\ N_{0}\ \Big(\frac{E}{E_{0}}\Big)^{-\Gamma}
\end{equation}
\begin{equation}\label{eq:ecpl}
\frac{dN}{dE}\ =\ N_{0}\ \Big(\frac{E}{E_{0}}\Big)^{-\Gamma}\ \exp\Big\lbrace-\frac{E}{E_{\rm C}}\Big\rbrace.
\end{equation}
\noindent{}$N_{0}$ is the normalization, $E_{0}$ = 0.445 GeV is the pivot energy from 3FGL, $\Gamma$ is the photon index, and $E_{\rm C}$ is the cutoff energy.  The best-fit spectral parameters are given in column 2 of Table \ref{tab:spectra}.  We detect a point source at the position of \mypsr\ with a likelihood test statistic \citep[][]{Mattox96} TS = 1967 (for one degree of freedom, the significance of a point source is $\sim\sqrt{\rm{TS}}$).

Using the likelihood ratio test, a single power-law shape is ruled out, in favor of an exponentially-cutoff power law, with a confidence level of $\sim$5.5$\sigma$, as indicated by the value of TS$_{\rm cut}$ \citep[defined as in][]{2PC}.  Table \ref{tab:spectra} also reports the photon ($F_{100}$) and energy ($G_{100}$) fluxes integrated from 0.1 to 100 GeV.  

We then analyzed $\sim$4.25 years of data up to 2012 November 12 (start of the estimated transition time window) and $\sim$2 years after 2012 December 21 (end of the estimated transition window) separately.  The results from the pre- and post-transition periods are given in columns 3 and 4 of Table \ref{tab:spectra}, respectively.  The photon flux above 100 MeV dropped by factor of $\sim$3 after the transition, the low-energy photon index hardened significantly, and the cutoff energy decreased.

\begin{figure}[h]
\epsscale{1.0}
\plotone{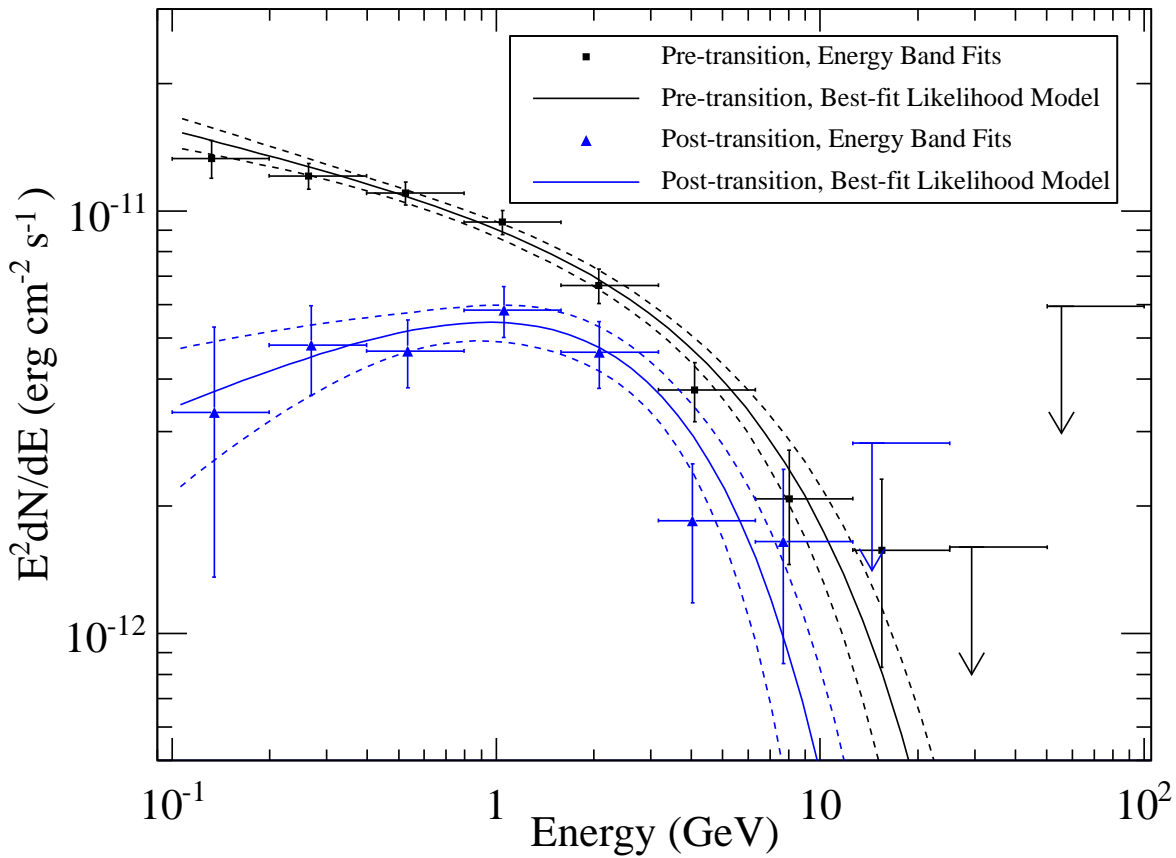}
\caption{Gamma-ray spectra of \mypsr\ pre-transition (black lines and square points) and post-transition (blue lines and triangular points).  The dashed lines show the 1$\sigma$ confidence region around the best-fit model for the pre- and post-transition data. \label{fig:spec}}
\end{figure}

For both the pre- and post-transition data sets, we calculated the spectral points in Figure \ref{fig:spec} by performing binned likelihood fits in the energy bands shown.  For each energy band, we started from the corresponding best-fit model and only freed the normalizations of sources within 6\DG\ of \mypsr\ and with TS $\geq$ 100 when fitting the full energy range.  We modeled the spectrum of \mypsr\ as a power law (Equation \ref{eq:pl}) with fixed photon index $\Gamma\ =$ 2.  For each energy band, we required that \mypsr\ be detected with TS $\geq$ 4 ($\sim2\sigma$) and with a predicted number of counts $\geq$ 4, else a 95\% confidence-level upper limit is plotted instead.  For each data set, we found the highest-energy event within the 95\% containment radius (as a function of event energy and angle with respect to the LAT boresight, $\theta$, and on the location of conversion in the LAT) and only fit up to the energy bin containing this energy, resulting in the pre-transition spectral measurements extending to higher energies.

\citet{Takata14} have proposed a model to explain the gamma-ray flux increase in the transition of MSP J1023+0038 \citep{Stappers14} in which the rotation-powered pulsar mechanism remains active (see Section \ref{models} for further discussion) and the increase in gamma-ray flux is due to the presence of a new, dominant component.  In order to investigate this possibility for \mypsr\ during the LMXB phase, we performed an additional binned maximum likelihood analysis of the pre-transition data set in which we had two point sources at the position of \mypsr.  The spectrum of the first source was set to the best-fit, exponentially-cutoff, power-law model from the post-transition data set with only the normalization parameter left free.  The second source was fit well by a simple power law with a photon index of $2.45\pm0.08$; the addition of an exponential cutoff was not statistically justified.  
There is no significant preference for the two-source fit over one source with a curved spectrum.  In the two-source fit, the normalization parameter of the first source did not change significantly relative to that found from fitting the post-transition data set.

\begin{deluxetable}{lccc}
\tablewidth{0pt}
\tablecaption{Spectral Fit Results \label{tab:spectra}}
\tablecolumns{4}
\tablehead{\colhead{Parameter} & \colhead{Full Data Set} & \colhead{Pre-transition} & \colhead{Post-transition}}
\startdata
$N_{0}$ (10$^{-11}$ cm$^{-2}$ s$^{-1}$ MeV$^{-1}$) & 3.19$\pm$0.12$^{+0.18}_{-0.16}$ & 3.78$\pm$0.16$^{+0.21}_{-0.20}$ & 1.86$\pm$0.23$^{+0.11}_{-0.09}$\\
$\Gamma$ & 2.13$\pm$0.05$^{+0.07}_{-0.06}$ & 2.18$\pm$0.06$^{+0.07}_{-0.06}$ & 1.66$\pm$0.22$\pm$0.06\\
$E_{\rm C}$ (GeV) & 6.8$\pm$1.6$^{+1.0}_{-0.7}$ & 7.5$\pm$2.2$^{+1.4}_{-0.9}$ & 2.8$\pm$1.1$\pm$0.2\\
$F_{100}$ (10$^{-8}$ cm$^{-2}$ s$^{-1}$) & 6.41$\pm$0.34$^{+0.44}_{-0.39}$ & 7.89$\pm$0.42$^{+0.57}_{-0.48}$ & 2.61$\pm$0.52$^{+0.15}_{-0.13}$\\
$G_{100}$ (10$^{-11}$ erg cm$^{-2}$ s$^{-1}$) & 3.50$\pm$0.12$^{+0.21}_{-0.19}$ & 4.16$\pm$0.15$^{+0.26}_{-0.23}$ & 1.86$\pm$0.19$^{+0.10}_{-0.09}$\\
TS & 1967 & 1745 & 266\\
TS$_{\rm cut}$ & 31 & 19 & 19\\
\enddata
\tablecomments{For all parameters, the first uncertainties are statistical, 1 $\sigma$, and the second reflect systematic uncertainties in the LAT effective area as detailed in the text.}
\end{deluxetable}
\clearpage

\subsubsection{SYSTEMATIC UNCERTAINTIES}\label{systs}
The systematic uncertainties in the LAT effective area ($A_{\rm{eff}}$) have been estimated to be 10\% for $\log_{10} (E/1\ \rm{MeV})\ \leq\ 2$, 5\% for $\log_{10} (E/1\ \rm{MeV})\ =\ 2.75$, and 10\% for $\log_{10} (E/1\ \rm{MeV})\ \geq\ 4$ with linear extrapolation, in log space, between these energies \citep{P7}.  Following \citet{2PC}, we estimated the effects of these uncertainties on our derived spectral parameters by generating bracketing IRFs using a modified $A_{\rm{eff}}$ given by,
\begin{equation}\label{eq:brack}
A_{\rm{brack}}(E,\theta)\ =\ A_{\rm{eff}}(E,\theta)(1+err(E)B(E)),
\end{equation}
\noindent where $err(E)$ represents the systematic uncertainties and we used $B(E)\ =\ \pm1$ as the bracketing function to estimate systematic uncertainties on $N_{0}$ and $B(E)\ =\ \pm\tanh(\log_{10}(E/E_{0})/0.13)$ for $\Gamma$ and $E_{\rm{C}}$, again with $E_{0}$ = 0.445 GeV.  For each fit with different bracketing IRFs, we generated new sourcemaps\footnote{For an explanation of LAT sourcemaps generation, see the LAT data analysis tutorial at\\ \url{http://fermi.gsfc.nasa.gov/ssc/data/analysis/scitools/binned_likelihood_tutorial.html\#computeSourceMaps}.} for point sources with free parameters but used the sourcemaps generated with the P7REP\_SOURCE\_V15 IRFs for the fixed sources and the diffuse components.  This last step is important as the diffuse components are tailored to the P7REP data and nominal IRFs, and the parameters from the fixed sources are from fits using the same IRFs.  To estimate the systematic uncertainties on $F_{100}$ and $G_{100}$, we recalculated these values from each bracketing IRF fit and took the maximum excursions from the nominal values as the systematic uncertainty.  These estimated systematic uncertainties are given as the second uncertainties on spectral parameters in Table \ref{tab:spectra}.


\subsection{LONG TERM SPECTRAL BEHAVIOR}\label{longterm}
Using a 15\DG\ radius selection we performed unbinned likelihood fits in 30 day time bins in order to characterize the gamma-ray flux change of \mypsr.  For each time bin we started from the best-fit model of the entire data set and modeled the spectrum of \mypsr\ as a simple power law (Equation \ref{eq:pl}) with both $N_{0}$ and $\Gamma$ free.  We kept the same point sources free as in the fit of the full data set, but only allowed the normalization parameters to vary.  In our first attempt, the time bin spanning 55702.66 MJD to 55732.66 MJD was found to have a flux above 100 MeV twice as high as the value obtained from fitting the pre-transition data alone.  No other 30-day time bins were found with similar flux values.  To further investigate this apparent high flux point, we used LAT ScienceTool \texttt{gttsmap} to make a 3\DG$\times$3\DG\ TS map, using unbinned likelihood, centered on the pulsar position in this 30-day time bin without the pulsar in the model.  The pulsar position is well outside the 99\% confidence-level contour of the TS map peak.  Using the LAT ScienceTool \texttt{gtfindsrc}, we localized this emission to right ascension $12^{\rm{h}}26^{\rm{m}}9\fs6$ and declination $-49^{\circ}35\arcmin24\arcsec$ (J2000) with a 95\% confidence-level radius of $r_{95}\ =\ 10\arcmin$, approximately $42\arcmin$ from the position of \mypsr.  Investigating this new source position, we found the blazar candidate PMN J1225$-$4936 \citep[CRATES J1225$-$4936,][]{Healey07} only $4\arcmin$ away, well within $r_{95}$.

A likelihood fit of the data for this 30-day time bin with \mypsr\ and PMN J1225$-$4936 using power-law spectra in the model detected both sources significantly, with the flux split approximately evenly between the two, thus bringing the flux of \mypsr\ in line with neighboring bins.  If we include the candidate blazar in the model of the region and fit over the entire data set, we find the source with TS = 4, explaining why there is no corresponding source in the 3FGL catalog.  A 30-day flux light curve analysis of the candidate blazar suggests that this is the only time bin in which the source is detected at $\geq$ 3$\sigma$ confidence level.  The candidate blazar was not found to be significant near the estimated transition time window. We computed similar TS maps in time bins when the candidate blazar was not significantly detected, during the pre-transition period, and found that the peak of the TS map was always consistent with the position of \mypsr\ within the 95\% confidence-level contour.  Therefore,  we conclude that PMN J1225$-$4936 is likely a transient gamma-ray blazar with one flare, reaching a photon flux of $(9.8\pm4.1)\times10^{-8}$ cm$^{-2}$ s$^{-1}$ with $\Gamma\ =\ 2.4\pm0.2$, during the reported time interval.

\begin{figure}[h]
\epsscale{1.0}
\plotone{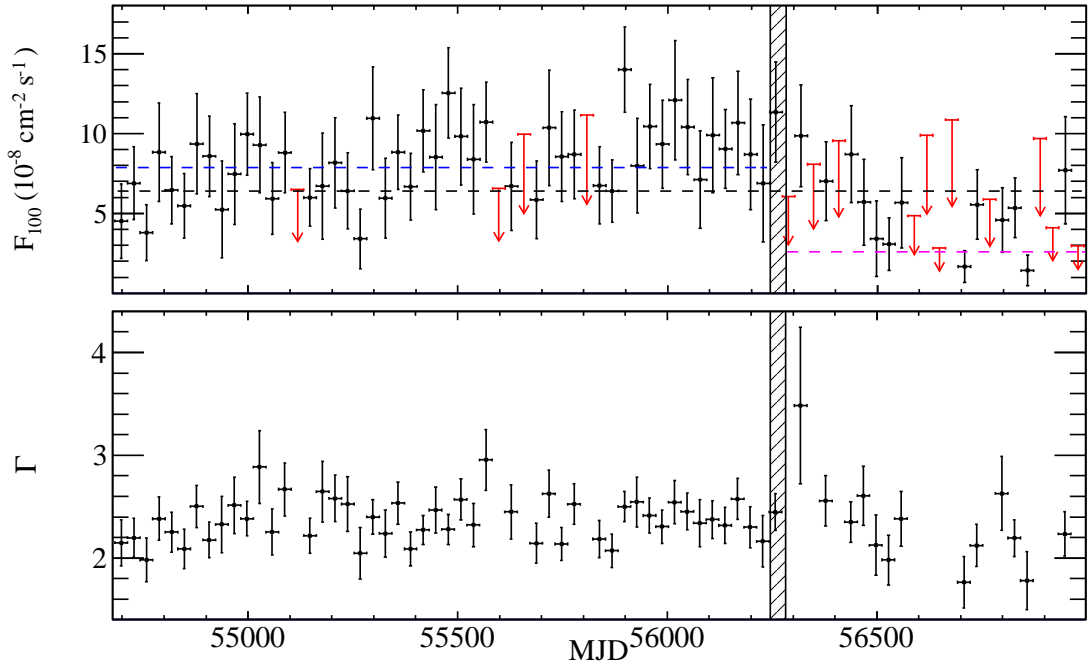}
\caption{The top panel shows $F_{100}$ in 30 day time bins.  Points are only shown if \mypsr\ was found with TS $\geq$ 12 ($\sim2\sigma$ for two degrees of freedom) and at least four predicted counts, else a 95\% confidence-level upper limit is shown (red); error bars are 1$\sigma$ statistical only from the likelihood fits.  The black dashed line shows the $F_{100}$ value from fitting the entire data set.  The blue dashed line shows the $F_{100}$ value from fitting only the pre-transition data.  The pink dashed line shows the $F_{100}$ value from fitting only the post-transition data.  The bottom panel shows the best-fit $\Gamma$ value for time bins where \mypsr\ was significantly detected.  In both panels, the diagonal hatched region shows the estimated transition time window from \citet{Bassa14}.  When maximizing the likelihood in each bin, a point source corresponding to the blazar candidate PMN J1225$-$4936 was included in the model of the region. \label{fig:fluxlc}}
\end{figure}

We then recomputed the 30-day flux light curve of \mypsr\ with PMN J1225$-$4936 included in the model with a fixed power-law index of 2.4.  Figure \ref{fig:fluxlc} shows $F_{100}$ versus time as well as the best-fit $\Gamma$ for those bins in which we do not report a flux upper limit.  We note that while the source was not found to be significantly variable during the first two years of the \fermi\ mission, it is flagged as variable in 3FGL; however, in Figure \ref{fig:fluxlc} the apparent variability is somewhat reduced compared to the result when the candidate blazar is not included in the model.  In order to assess if the source should still be considered significantly variable once the candidate blazar is included in the model, we computed a spectral variability index for the pre-transition data similar to the flux variability index used in the \fermi\ LAT catalogs \citep{2FGL}. To do this, we repeated the spectral analysis in each 30-day time bin with the spectrum of \mypsr\ fixed to the best-fit power-law model over the pre-transition data, recorded the negative log likelihood values ($-\ln \mathcal{L}_{\rm{null},i}$) and compared these to the values from the fits with the pulsar spectral parameters free ($-\ln \mathcal{L}_{i}$) to construct the spectral variability index as,
\begin{equation}\label{eq:vi}
TS_{\rm var}\ =\ -2\Big(\sum_{i}[(-\ln \mathcal{L}_{i} ) - (-\ln \mathcal{L}_{\rm{null},i} )]\Big).
\end{equation}

Assuming that the variability index is distributed as a $\chi^{2}$ with 50 degrees of freedom (52 months minus 2 since we let both the index and normalization be free in the fits) we need $TS_{\rm var}\ \gtrsim$ 76.15 to say the source is variable to better than 99\% confidence.  Using Equation \ref{eq:vi} we find $TS_{\rm var}\ =$ 95, compared to $TS_{\rm var}\ =$ 125 without the candidate blazar in the model, suggesting that the spectrum is still variable during the pre-transition period but the amount of variability is reduced.
It is important to note that \fermi\ LAT catalogs only allow the source normalization to be free when computing the variability index, whereas we let both the index and normalization be free, so while the catalog analysis addresses whether the flux is variable our analysis addresses whether the entire spectrum is variable.


The LAT upper limit approximately halfway through the transition time window indicated by the X-ray observations \citep{Bassa14} could provide a more precise estimate of the transition epoch.  Examination of a flux light curve in two-day bins (not shown) bracketing the transition time window indicates a significant drop in TS near 2012 November 30 with predominantly TS $\geq$ 4 (as high as 26) before and TS $<$ 4 (with one bin having TS = 5) after.  We have verified that the drop in TS after this date is not due to decreased exposure.  However, the flux appears to settle gradually in the post-transition time period with the emission starting softer than the pre-transition emission and gradually hardening.  This may suggest that the transition from LMXB to rotation-powered pulsar state progresses over a longer time period than when transitioning in the other direction, as implied by the abrupt increase in gamma-ray flux seen for PSR J1023+0038 \citep{Stappers14}.

\subsection{PULSED GAMMA RAYS}\label{pulses}
To test for gamma-ray pulsations from \mypsr\ we kept only events within a 3\DG\ radius of the radio position after the transition and assigned a rotational phase to each LAT event using the timing solution of \citet{J1227Radio} \and the \texttt{fermi} plugin \citep{Ray11} to the  \textsc{Tempo2}\footnote{\url{http://sourceforge.net/projects/tempo2/}} software \citep{T2}.  We tested for pulsations in the LAT data from the  start of radio monitoring (MJD 56707.98) to the end of the data set described in Section \ref{prep}. Using the H test \citep{deJ89,deJ10} together with weights for each photon derived from the spectral analysis \citep[as formulated by][]{Kerr11}, we obtain a test statistic value H = 35.3, implying a chance probability of 7.3$\times10^{-7}$, or a 5.0$\sigma$ significance, indicating a significant detection of gamma-ray pulsations.  The H test is a powerful test for rejecting the null hypothesis of a uniform distribution of phases, and including the photon weights, which represents the probability a photon originated from the pulsar instead of a different source, further increases its sensitivity.  For in-depth descriptions of the test and its applicability to gamma-ray pulsar data see \citet{deJ89,deJ10,Kerr11,2PC}.  As a check on our detection, we assigned random phases, from a uniform distribution between 0 and 1, to our calculated spectral weights and recalculated the H statistic.  In 1$\times10^{5}$ trials, the largest H-test value we found was 28.5, indicating that the probability of obtaining an H-test value $>$ 28.5 by chance was $<\ 1\times10^{-5}$, supporting the significance of our detection.

We investigated how far backward in time, toward the end of the estimated transition time window, we could extend our data set for the gamma-ray light curve.  The pulsed significance continued to increase when adding data back to approximately MJD 56650 ($\sim$2 months before the start of the ephemeris validity range).  Adding events before this date made the detection less significant, indicating that the timing model is not sufficient to extrapolate backward beyond this point.  This is consistent with the fact that the radio model determined from the 2014 February through December timing data does not extrapolate backward to the 2013 November data, as found by \citet{J1227Radio}.

We also explored the possibility of extending the pulsar ephemeris back in time before the radio observations using the extra year of LAT gamma-ray data after the transition. An extended timing solution could lead to better constraints on the system, especially the orbital parameters, owing to the longer baseline and would address the question of whether the system has been an active gamma-ray pulsar since the state transition. However, the relatively low signal-to-noise ratio of the gamma-ray pulsations from \mypsr\ precludes derivation of a reliable timing solution with the LAT. In particular, exploring the parameter space in the neighborhood of the radio ephemeris resulted in many local maxima, but each with large parameter uncertainties and too low statistical significance to be confidently trusted.  Given the potential insights to be gained from being able to extrapolate the timing solution back further, this is certainly worth revisiting with more data and the forthcoming, more sensitive Pass 8 LAT data \citep{Pass8}.

\begin{figure}[h]
\epsscale{1.0}
\plotone{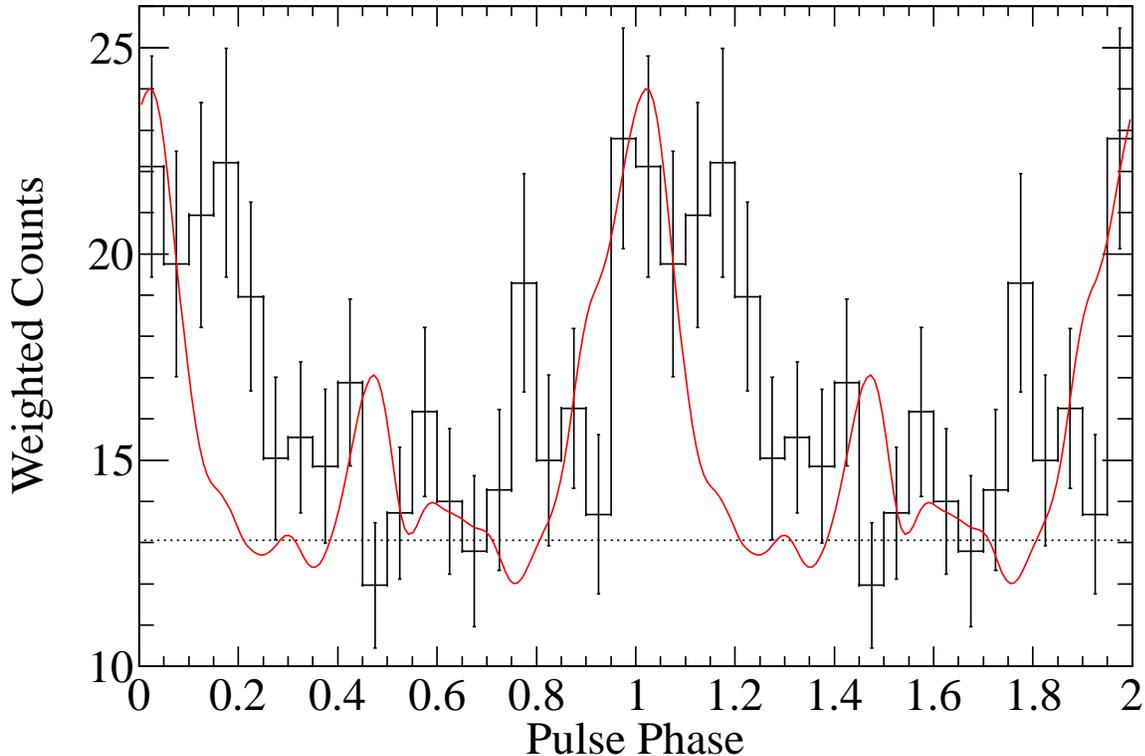}
\caption{Phase-aligned gamma-ray (histogram, 20 bins per rotation) and 1.4 GHz Parkes (red line) light curves of \mypsr, with two rotations shown for clarity; the radio profile has been renormalized and the zero level adjusted for ease of viewing and has arbitrary units.  The statistical uncertainties for the LAT light curve bins and the gamma-ray background level, horizontal dotted line, are derived from the spectral weights estimated as in \citet{2PC}.  The low-level peak at phase $\sim$0.5 in the Parkes light curve is a real interpulse, which becomes dominant at lower frequencies. \label{fig:lc}}
\end{figure}

The $\geq$ 100 MeV light curve of \mypsr\ using all events within a 3\DG\ radius of the radio position and detected between 56650 MJD and the end of our data set is shown in Figure \ref{fig:lc}.  The Parkes 1.4 GHz radio profile from \citet{J1227Radio} is over-plotted.  The absolute phase alignment was done by using \textsc{Tempo2} to define pulse phase 0 using the 1.4 GHz radio TOAs, and plotting the Parkes radio template with the LAT light curve aligned to that fiducial phase using the \texttt{fermi} plugin for \textsc{Tempo2}.
\newpage

\begin{figure}[h]
\epsscale{1.0}
\plotone{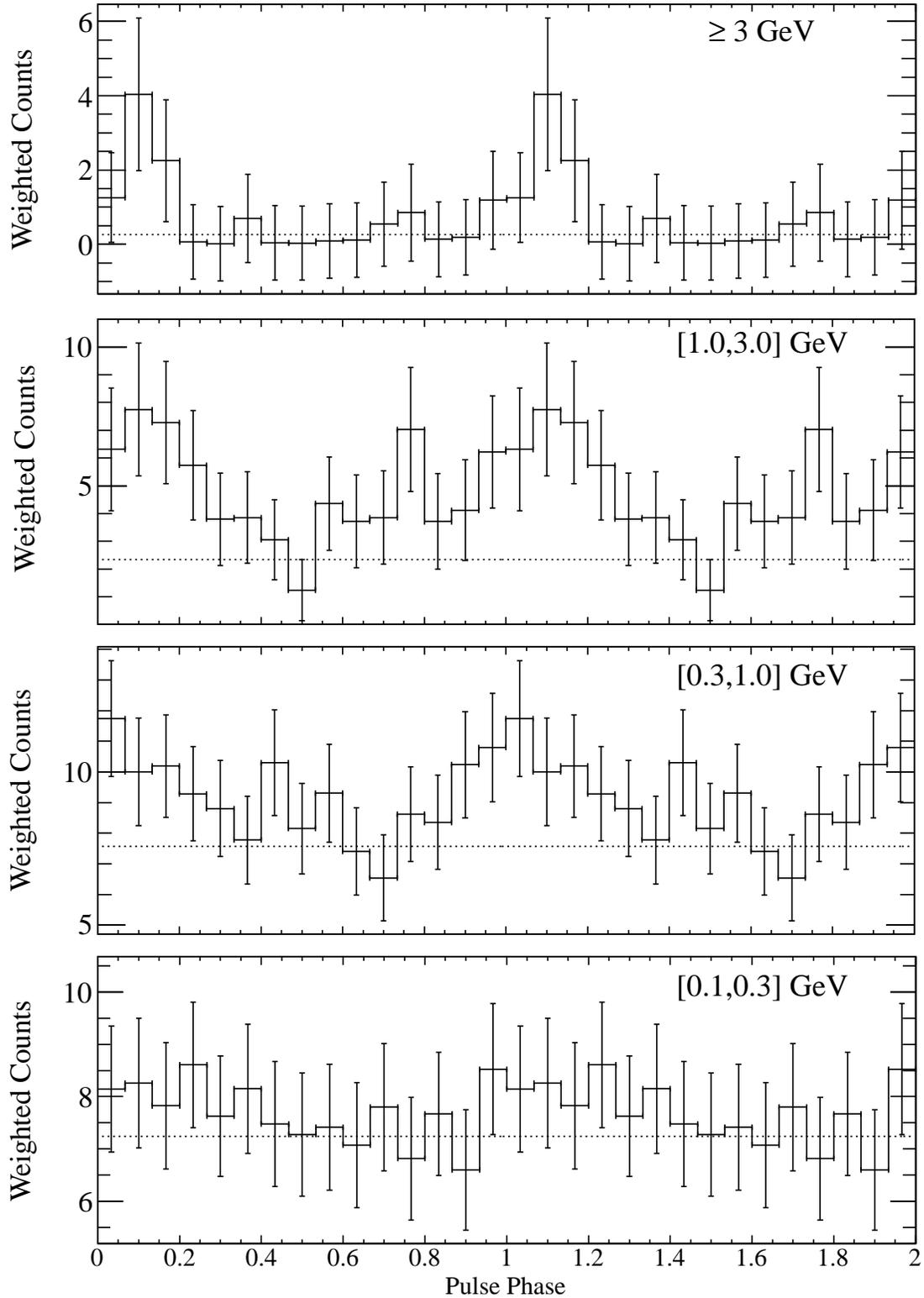}
\caption{Gamma-ray light curves of \mypsr\ (15 bins per rotation) in 4 energy bands as labeled, with two rotations shown for clarity.  The horizontal dotted lines in each panel show the estimated background levels.  Uncertainties and background levels are estimated as for Figure \ref{fig:lc}. \label{fig:enelc}}
\end{figure}

The gamma-ray light curve of \mypsr\ can be satisfactorily fit with one broad Gaussian, FWHM = $0.38\pm0.09$, at $0.09\pm0.03$ in phase (assuming the estimated background level from the weights to reduce the number of fit parameters).  A fit with two closely-spaced Gaussian peaks is also acceptable, but the improvement over the single-peak fit is not significant with the current statistics.  Figure \ref{fig:enelc} shows the energy evolution of the gamma-ray light curve.  When only considering events with energies from 0.3 to 1.0 GeV, the light curve peak is almost at phase 0 while when looking at the light curve for events with energies $\geq$ 3 GeV the peak is sharper and near 0.1 in phase.  This behavior is similar to what is seen for other pulsars with two closely-spaced peaks \citep[e.g., PSR J0007+7303,][]{2PC}.

\clearpage

The gamma-ray peak is nearly aligned with the main peak in the 1.4 GHz Parkes radio profile.  \citet{J1227Radio} also presented the pulse profile of \mypsr\ at 607 and 322 MHz.  The lower-level peak in the 1.4 GHz profile, near $\sim$0.5 in phase, is a real feature and becomes dominant at lower frequencies.  In fact, at 322 MHz the main 1.4 GHz peak disappears entirely.

\subsection{TEST FOR ORBITAL MODULATION}\label{orbit}
In order to test for modulation at the orbital period, we first made a counts light curve in 30 second time bins from the post-transition data set and then calculated the orbital phase and LAT exposure for each time bin.  This allows us to correct for potential exposure variations across the orbital period, which has been shown to be a necessary step to avoid detection of false modulation \citep{KerrThesis,J1018}.  



We created a new null distribution ($\Phi_{\rm null}$) for the variability tests by binning the exposure in 1000 bins of orbital phase and normalizing.  This $\Phi_{\rm null}$ reflects the variation with orbital phase we would expect for a non-varying source due to differences in exposure. Following \citet{KerrThesis}, we used our new $\Phi_{\rm null}$ and the spectral weights calculated in Section \ref{pulses} in a weighted H test and Z$_{m}^{2}$ test with two harmonics to test for modulation at the orbital period.  The significance values from the weighted H and Z$_{m}^{2}$ tests were 1.1$\sigma$ and 1.4$\sigma$, respectively, indicating no strong evidence for modulation at the orbital period.

\begin{figure}[h]
\epsscale{1.0}
\plotone{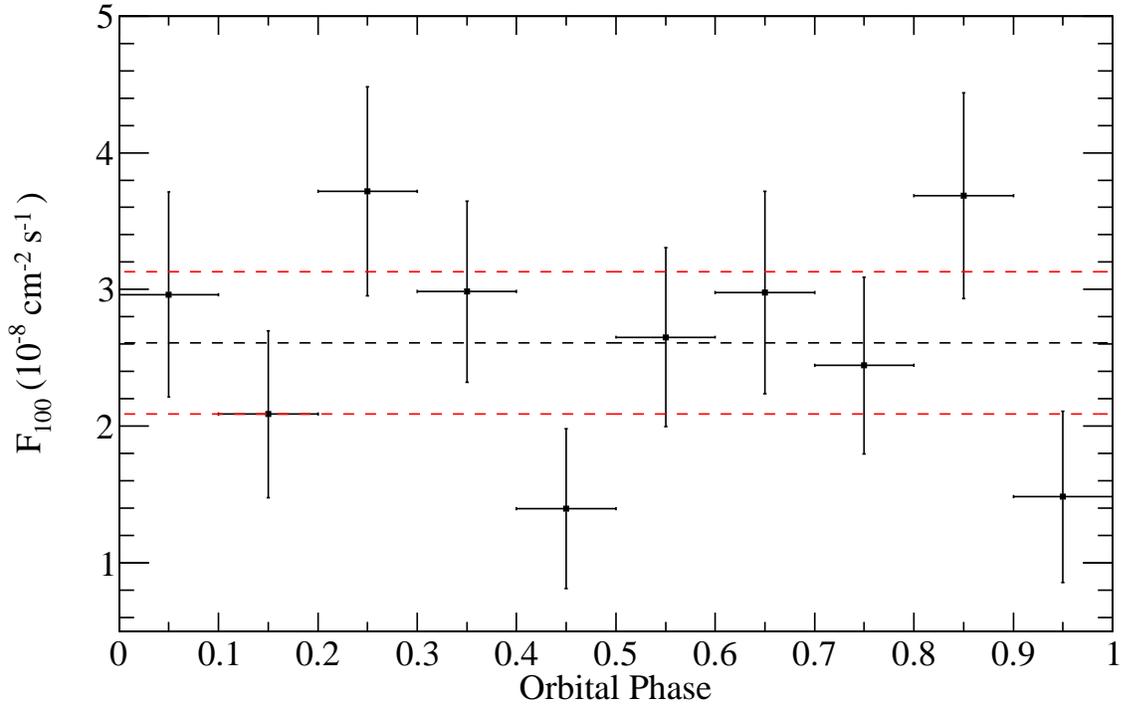}
\caption{Flux light curve versus orbital phase for \mypsr\ for the full post-transition data set; uncertainties are 1$\sigma$ statistical only from the likelihood fits.  Each point was found with a significance $\geq$ 2$\sigma$.  The dashed black line is the phase-averaged flux given in Table \ref{tab:spectra} and the dashed red lines show $\pm1\sigma$ (statistical only).\label{fig:olc}}
\end{figure}

We split the post-transition data into 10 orbital phase bins, using the binary period from the radio timing solution.   The good time intervals for each bin were corrected to properly account for the orbital phase selections.  Starting from the best-fit model with the spectrum of \mypsr\ modeled as an exponentially cutoff power law (Equation \ref{eq:ecpl}), we kept only the normalization parameters of sources free and performed binned likelihood fits in each bin.  The resulting orbital flux light curve is shown in Figure \ref{fig:olc}.

Using the phase-averaged flux in a $\chi^{2}$ analysis (neglecting the uncertainty on the phase-averaged value) gives a reduced $\chi^{2}$ value of 1.47 (with 9 degrees of freedom), a larger $\chi^{2}$ value would be found 15\% of the time by chance, suggesting this is an acceptable fit.  This reduced $\chi^{2}$ value suggests that any variations in Figure \ref{fig:olc} are consistent with statistical fluctuations, supporting the conclusion that the gamma-ray emission from \mypsr\ is not modulated at the orbital period.

\section{DISCUSSION}\label{disc}
\citet{Hill11} used $\sim$2 years of LAT \texttt{Pass 6} data to analyze the region around \mypsr\ while it was still, we now recognize, in the LMXB phase.  Our best-fit $\Gamma$ and derived photon flux for the pre-transition data set agree with their results, but our best-fit cutoff energy is nearly twice their value of $E_{\rm C}\ =\ 4.1\pm1.3$ GeV, though consistent within the joint uncertainties.

\citet{XW14} present a similar LAT study of \mypsr\ using $\sim$6 years of data.  Their spectral analysis was restricted to events with energies $\geq$ 0.2 GeV.  Using our best fit, we derive a flux from 0.2 to 100 GeV of $(2.79\pm0.96)\times10^{-8}$ cm$^{-2}$ s$^{-1}$ over the entire data set, statistical uncertainty only; this value and our best-fit $\Gamma$ and $E_{\rm C}$ are in good agreement with their results.  Additionally, our best-fit, pre-transition model gives a flux from 0.2 to 100 GeV of $(3.34\pm0.12)\times10^{-8}$ cm$^{-2}$ s$^{-1}$, and our best post-fit transition model gives $(1.44\pm0.17)\times10^{-8}$ cm$^{-2}$ s$^{-1}$, statistical uncertainties only.  Both of these values and the other spectral parameters are in good agreement with their results.

A direct comparison of our flux light curve to that of \citet{XW14} is complicated by their use of 0.2 GeV as a minimum energy and by the fact that they do not impose a cut on TS when deciding whether to plot points or upper limits in the flux light curve.  The spectrum is still variable during the pre-transition data but the significance of the variability is reduced in our result due to the inclusion of the source we associate with PMN J1225$-$4936.  The $\geq$ 0.2 GeV flux light curve of \citet{XW14} does show a similar increase around the time of our claimed flare from this candidate blazar.

\citet{XW14} have claimed marginal evidence for modulation of the gamma-ray emission from \mypsr\ at the optical period of \citet{Bassa14} after the transition, which is not seen in our analysis.  Using the optical period and 10 orbital phase bins, we found that the exposure varies by $\lesssim$ 2\% from the average in any given bin, indicating that their claimed orbital modulation is not likely due to varying exposure alone.  We note, however, that their stated H test value of $\sim$10 corresponds to a 2.4$\sigma$ detection, not 3$\sigma$ as claimed, and trials factors must be accounted for, reducing the significance even further.  In attempting to verify this claim we folded the same time interval of LAT data used in their paper at the optical period and used the same energy and radius cuts, but could only recover an unweighted H-test value of 8.1 (2.1$\sigma$).  Our orbital flux light curve (Figure \ref{fig:olc}) does have two bins that are 1.5--2$\sigma$ below the phase-averaged flux value, but both are well away from superior conjunction ($\phi_{\rm orb}\ =\ 0.25$) and thus are not easily explained as eclipses of the gamma-ray emission site by the companion \citep[a proposed explanation for the dip seen in the orbital light curve of][]{XW14}.

\subsection{ENERGETICS}\label{energ}
An important quantity for understanding the pulsar emission mechanism is the gamma-ray luminosity, defined as $L_{\gamma}\ =\ 4\pi f_{\Omega} G_{100} d^2$, where $f_{\Omega}$ is a beaming correction factor accounting for the non-isotropic nature of the emission \citep[e.g.,][]{Watters09} and $d$ is the distance to the pulsar.  The radio timing solution gives a dispersion measure (DM) of 43.4 pc cm$^{-3}$ for \mypsr\ \citep{J1227Radio}.  Using the NE2001 electron-density distribution \citep{NE2001} yields $d\ =\ 1.4\pm0.2$ kpc\footnote{The distance uncertainty is calculated by assuming a $\pm$20\% uncertainty on the DM, which is much larger than the formal DM uncertainty from the timing solution and is meant to reflect uncertainties in the electron-density distribution model, and calculating the resulting change in $d$ using the online NE2001 implementation \url{http://www.nrl.navy.mil/rsd/RORF/ne2001/}.}.

\citet{Mod2PC} fit the gamma-ray and radio light curves of all MSPs in the second LAT catalog of gamma-ray pulsars \citep{2PC}, using models that assumed the bulk of the gamma-ray emission comes from regions high in the pulsar magnetosphere.  They found that, for configurations capable of producing light curves that match the observations, $f_{\Omega}$ values are typically close to but slightly less than 1.  For the remainder of the discussion we will assume $f_{\Omega}\ =\ 1$.

Using the post-transition spectral results and the DM distance, we find $L_{\gamma}\ =\ (4.4\pm0.5\pm1.3)\times 10^{33}$ erg s$^{-1}$ \citep[following][we separate the uncertainty due to the gamma-ray spectral fit, first value, and that due to the distance, second value]{2PC}.  We define the gamma-ray efficiency as $\eta_{\gamma}\ =\ (L_{\gamma}/\dot{E}) \times 100$\%, where $\dot{E}\ =\ 4\pi^{2} I \dot{P}/P^{3}\ =\ 0.9\times10^{35}$ erg s$^{-1}$ is the spin-down luminosity of \mypsr\ \citep{J1227Radio}, $I$ is the moment of inertia and is taken to be $10^{45}$ g cm$^{2}$, \Per\ is the spin period of the pulsar in seconds, and \Pd\ is the time derivative of the pulsar spin period.  Using this definition and the values for \mypsr\ yields $\eta_{\gamma}\ =\ (4.9\pm0.6\pm1.4)$\%.  Comparing to other gamma-ray pulsars with similar \Ed, an efficiency of $\sim$5\% is on the low end but consistent with the $L_{\gamma}$ vs. \Ed\ trend seen for LAT pulsars \citep{2PC}.

It is also of interest to examine the energetics of the pre-transition emission.  Similar considerations give $L_{\gamma}\ =\ (9.8\pm0.4\pm2.8)\times 10^{33}$ erg s$^{-1}$ and $\eta_{\gamma}\ =\ (10.9\pm0.4\pm3.1)$\%, indicating that there is sufficient rotational energy to power that emission as well.  

\subsection{EMISSION MODELS}\label{models}
\subsubsection{PULSED LIGHT CURVES}\label{lcmodels}
The prevailing models for high-energy pulsar gamma-ray emission, such as the outer gap \citep[OG, e.g.,][]{OG} and slot gap \citep[SG,][]{MH03,MH04} models, assume that particles are accelerated in vacuum gaps along curved magnetic field lines and emit photons in the outer magnetosphere, far from the stellar surface.  
In the OG model, particles are assumed to be accelerated and emit gamma rays via curvature radiation in a vacuum gap above the null-charge surface (the geometric surface defined by the condition $\vec{\Omega}\cdot\vec{B}$ = 0), bordered by the last-closed field lines (those which close at the light cylinder and are able to rotate with the pulsar).  The SG model, on the other hand, assumes particles are accelerated in gaps bordering the last-closed field lines, emitting gamma rays along the way, from the stellar surface to near the light cylinder.  In both models, the bright, and typically sharp, peaks observed in gamma-ray pulsar light curves are caustics where time-of-flight delays and relativistic aberration result in photons emitted at different altitudes on trailing field lines arriving at an observer coincident in phase \citep[e.g.,][]{Morini83,Venter09}.  For both the SG and OG models, the vacuum gaps are assumed to be relatively narrow, a few percent of the polar cap size, as the rest of the magnetosphere is thought to be filled with charges that short out the accelerating electric field.  The pair-starved polar cap model \citep[PSPC, e.g.,][]{Harding05}, however, was developed for MSPs and old non-recycled pulsars where it was thought that the weaker magnetic fields would not be capable of producing enough charges, through pair cascades, to screen the accelerating field over the entire polar cap.  This means that the entire open field line region above the polar cap is available to accelerate particles and produce gamma rays.

With standard, low-altitude models of pulsar radio emission, the OG, SG, and PSPC models predict significant phase lags between features in the radio and gamma-ray light curves, observed for all non-recycled pulsars (except the Crab) and most MSPs known to emit gamma rays \citep{2PC}.  However, a subclass of gamma-ray MSPs has been established in which the gamma-ray and radio peaks occur at (nearly) the same rotational phase.  One explanation for this subclass has been that the radio emission is also produced far from the magnetosphere, co-located with the gamma-ray emission region, described by altitude-limited versions of the the OG and SG models \citep[e.g.,][]{Venter12,Mod2PC}.  \citet{Venter12} also introduced the low-altitude slot gap (laSG) model for this subclass of gamma-ray MSPs.  In the laSG model, the gamma rays and radio are both produced within a few stellar radii of the neutron star surface in a narrow gap near the last-closed field lines.  
Similar to PSR J1824$-$2452A \citep[in the globular cluster M28,][]{Johnson1821} the radio and gamma-ray light curves of \mypsr\ appear to have some aligned and some non-aligned features, so \mypsr\ does not cleanly fit into this gamma-ray MSP subclass, though the short spin period and large \Ed\ would make it, \emph{a priori}, a prime candidate.

\citet{Mod2PC} jointly fit the radio and gamma-ray light curves of MSPs in the second LAT catalog of gamma-ray pulsars using geometric realizations of the OG, two-pole caustic \citep[TPC,][taken to be a geometric realization of the SG]{TPC}, PSPC, and laSG gamma-ray emission models and hollow-cone plus core beam \citep[e.g.,][]{Story07}, altitude-limited versions of the OG and TPC models, or laSG radio emission models.  With the current statistics for \mypsr, a constraining fit is not possible, but qualitative comments can be made concerning the probable system geometry and viable emission models.  The emission geometry of a pulsar is typically defined by two angles, the angle between the spin and magnetic axes ($\alpha$) and the angle between the magnetic axis and the observer's line of sight ($\beta$).  Often, it is instead desirable to work in terms of $\alpha$ and the angle between the spin axis and the observer's line of sight ($\zeta\ \equiv\ \beta-\alpha$).  For pulsars in binary systems, another important angle is that between the orbital axis and the observer's line of sight ($i$).

\citet{deM14} used optical observations of XSS J12270$-$4859 to constrain the orbital inclination to be  45\DG\ $\lesssim\ i\ \lesssim$ 65\DG.  The recycling process is thought to align the orbit and spin axes, such that we can assume $i\ \sim\ \zeta$, translating the previous constraints into limits on $\zeta$.  Similarly, the observation of radio eclipses for significant portions of the orbit near superior conjunction \citep{J1227Radio} suggest $i\ \gtrsim$ 60\DG.  The radio profile at 1.4 GHz (Figure \ref{fig:lc}) consists of two peaks separated by approximately 0.5 in phase.  Assuming a hollow-cone beam and $\zeta$ consistent with the optical constraints on $i$, the radio profile suggests $\alpha$ near 90\DG, but not exactly 90\DG\ as the two peaks are not symmetric.  However, standard outer-magnetospheric gamma-ray emission models cannot produce the observed peak separation and phase lag with respect to the radio for those geometries.  Acceptable simulated gamma-ray light curves can be found using the OG, TPC, or PSPC models and a hollow-cone radio beam with reasonable values of $\zeta$, but all require low values of $\alpha$ ($\lesssim$ 20\DG) and predict only one radio peak.  If we fit only the gamma-ray light curves, the TPC model finds a best-fit geometry with $\zeta$ near 65\DG, agreeing well with constraints on $i$, and $\alpha$ near 30\DG, while the OG model prefers large $\alpha$ values near 80\DG\ and small $\zeta$ near 15\DG, not in agreement with the constraints on $i$.  We remind the reader, however, that with the current gamma-ray statistics it is difficult to\newpage

\noindent{}make definitive statements regarding emission geometries that can reproduce the observed light curves of \mypsr.


Another possibility is that the 1.4 GHz radio peak that is nearly aligned with the gamma-ray peak originates in an extended region in the outer magnetosphere co-located with the gamma-ray emission region, while the other radio peak arises from lower altitude nearer the polar cap. This model would eliminate the need for $\alpha$ to be near 90\DG\ in order for two conventional radio peaks to be visible.  The gamma-ray only TPC geometry with $\alpha$ near 30\DG\ and $\zeta$ near 65\DG\ would then be consistent with the orbital inclination constraint and with single gamma-ray and radio caustic peaks.  Different locations for the two radio peaks could also account for their very different spectral evolution.  Radio polarization measurements could verify this model, which would predict depolarization and rapid position angle swings in the caustic peaks but higher levels of polarization in the other peak.

\citet{AnnGap1821} reproduced the 1.4 GHz radio, X-ray, and gamma-ray light curves of PSR J1824$-$2452A using a single-pole annular gap model.  The emission regions for all three wavebands were modeled as extended in the pulsar magnetosphere with some overlap, explaining the fact that some, but not all, of the peaks are aligned in phase.  A similar model may be able to explain the radio and gamma-ray light curves of \mypsr, though an independent constraint on $\alpha$ would be helpful for such an endeavor as those authors have not yet developed a statistical technique to choose the best-fit geometry from profile fits.

\subsubsection{PRE-TRANSITION EMISSION}\label{premodels}
\citet{Takata14} analyzed multi-wavelength data for PSR J1023+0038, characterizing the transition of the system from rotation to accretion-powered state that occurred near the end of 2013 June.  They proposed an emission model in which an accretion disk forms but does not penetrate inside the light cylinder (defined by the cylindrical radius at which rotation with the star requires moving at the speed of light) suggesting that the rotation-powered pulsar emission mechanisms are still active.  In their model, the observed gamma-ray flux increase represents the emergence of an additional component, not modulated at the orbital period, due to inverse Compton scattering of the pulsar wind off of ultraviolet photons from the accretion disk.  From the two-source fit of the \mypsr\ pre-transition data detailed in Section \ref{spec}, we cannot rule out that the rotation-powered pulsar mechanism was active while the source was in the LMXB state.  However, without a detection of gamma-ray pulsations during this time period we cannot confidently say that it was active either.

\citet{J1227XMM} argued that the X-ray pulsations at the spin period during the LMXB phase from \mypsr\ are most plausibly explained as accretion rather than rotation-powered, which argues against the model of \citet{Takata14} to explain the pre-transition gamma-ray emission.  Recent observations of PSR J1023+0038 during the LMXB state also argue against this model.  \citet{Archibald14} have detected coherent X-ray pulsations, most likely accretion powered, from PSR J1023+0038 after the recent state transition.  Radio observations, at frequencies up to $\sim$5 GHz, have failed to detect pulsations at the rotational period while the source is in the LMXB state \citep{BogdanovJ1023Xray}.  Finally, radio imaging observations by \citet{DellerJ1023Radio}, during the LMXB state, found a variable, flat-spectrum radio source that is incompatible with the previously very-steep spectrum of PSR J1023+0038.

\citet{PTL14} proposed a propellor model to explain the pre-transition (LMXB phase) X- and gamma-ray emission from \mypsr.  In their model, the accretion disk penetrates inside the light cylinder (with an inner disk radius of $\sim$40 km for \mypsr) quenching the rotation-powered pulsar emission mechanism.  The observed X- and gamma-ray emission is the result of synchrotron and synchrotron self-Compton scattering from electrons accelerated at the interface between the disk and the pulsar magnetosphere.  For reasonable parameters, their model was able to reproduce the spectrum of \citet{Hill11}.  Both the spin period and inferred dipole magnetic field strength reported by \citet{J1227Radio} are in good agreement with the values assumed by \citet{PTL14}, suggesting that their model is a viable explanation of the emission during the LMXB phase.  \citet{J1227XMM} inferred an accretion disk truncation radius for XSS J12270$-$4859, from the luminosity of the X-ray pulsations, that was outside of the corotation radius, which would prevent accretion.  To reconcile this with the apparent accretion-powered nature of their observed X-ray pulsations, they theorized that the disk mass accretion rate was actually $\sim$30 times larger than their inferred value, but that only a small portion was channeled to the neutron star surface with the rest being ejected in an outflow, a scenario that is in agreement with the radio spectrum observed by \citet{Hill11}.  Such an outflow would argue in favor of the propellor model.

\section{CONCLUSIONS}\label{conc}
It is now clear that \mypsr\ is one of three redback MSPs, to date, observed to have transitioned between an LMXB state and a rotation-powered pulsar.  We have detected significant gamma-ray pulsations from \mypsr\ after the transition in late 2012 and characterized the spectral differences pre- and post-transition.  Analysis of the LAT data during the transition time window in 2-day time bins suggests that the transition occurred near 2012 November 30.  After the transition the photon (energy) flux dropped by a factor of $\sim$3 ($\sim$2), the photon index hardened, and the cutoff energy decreased.  We find no evidence in the LAT data, pre- and post-transition, for significant modulation at the orbital period.  PSR J1023+0038 has been observed to transition in the other direction, from rotation-powered pulsar to LMXB, displaying a larger gamma-ray flux increase, with estimates of the change varying from a factor of $\sim$5 \citep{Stappers14} to $\sim$10 \citep{Takata14}.  PSR J1023+0038 also displays a lower cutoff energy and harder photon index during the rotation-powered state compared to the LMXB state \citep{Takata14}.  During the rotation-powered pulsar state, the gamma-ray flux from \mypsr\ is $\sim$4 times greater than that from PSR J1023+0038, which is at a similar distance (1.4 kpc) but has a lower \Ed\ (4.4$\times10^{34}$ erg s$^{-1}$) \citep{J1023timing}.  While the viewing geometry is not precisely known for either system, we can speculate that the difference in pulsed flux is likely due to beaming and a less favorable geometry for PSR J1023+0038 \citep[e.g., Fig.~20 of][]{Venter09}.  During the LMXB state the observed gamma-ray fluxes from \mypsr\ and PSR J1023+0038 are very similar, but \mypsr\ has a softer photon index and higher cutoff energy in this state than PSR J1023+0038.  

The gamma-ray light curve can be reproduced with outer-magnetospheric emission models for geometries that agree with the observation of radio eclipses and optical constraints on $i$, assuming that the orbital and spin axes have aligned through the recycling process.  The near alignment of the gamma-ray peak with the main 1.4 GHz radio peak suggests that at least part of the radio emission may be generated in extended regions at high altitude in the magnetosphere, co-located with the gamma-ray emission site.  With PSR J1824$-$2452A, this is the second gamma-ray MSP with high spin-down power ($\gtrsim10^{35}$ erg s$^{-1}$) but only partially aligned light curve peaks, as opposed to PSRs J1823$-$3021A and B1937+21 for which the gamma-ray peaks are all aligned with radio features \citep[e.g.,][]{Mod2PC}.  PSR B1957+20 represents a mix between the MSP subclass with aligned radio and gamma-ray peaks and the partially aligned MSPs.  In particular, \citet{Guillemot12} showed that at 0.35 GHz, both radio and gamma-ray peaks occur at the same rotational phase; however, an additional radio peak is present at 1.4 GHz that is not matched by a feature in the gamma-ray light curve.  \cite{J1023timing} reported marginal evidence (3.7$\sigma$) for gamma-ray pulsations from PSR J1023+0038, with a suggested gamma-ray peak significantly separated in phase from the radio peaks (by 0.4 to 0.5 in phase).   There is no evidence for modulation of the gamma-ray flux from \mypsr\ at the orbital period during the pre- or post-transition periods.  Continued study of the pulsed emission of \mypsr\, such as increased gamma-ray statistics, radio polarization, and searches for pulsations at other wavelengths, should provide insights and help to further test and refine pulsar emission models.

Coincident changes in the gamma-ray flux accompanied two of the three transitions in redback systems observed to date.  In the case of PSR J1824$-$2452I, a non-detection is not particularly troubling given the distance \citep[5.1$\pm$0.5 kpc,][]{RC91} and the added complication of a bright foreground signal in the LAT data from PSR J1824$-$2452A.  As such, it is of interest to speculate on the rate at which redback systems undergo such transitions and whether we should expect to observe more transitions with the LAT.  We know of nine redback systems in the Galactic field, which we can use to estimate a lower limit (acknowledging that there are unknown redbacks among the unassociated LAT sources) to the number of redback-years that LAT has observed to be 54.  With two events observed during the \fermi\ mission, we can estimate the rate at which a redback undergoes a transition to be once every $\gtrsim$ 27 years.  We can check this estimate against the fact that PSR J1023+0038 showed evidence for an accretion disk in the 2000--2001 time frame, was detected as a radio pulsar in 2009, and transitioned back to an accreting state in 2013 \citep{Archibald09,Stappers14}.  This suggests a timescale on the order of a decade between transitions, in rough agreement with our estimated lower limit.  This lower limit on the rate may in fact be overestimated if we consider that not all of the known redback systems in the Galactic field should be expected to transition at all.  It is currently unclear what triggers a transition, in either direction.  PSRs J1227$-$4853 and J1023+0038 both have \Per\ $\approx$ 1.69 ms and orbital periods less than 7 hours.  If we only include redbacks with short spin ($<$ 3 ms) and orbital periods ($<$ 7 hours) we have four sources and our estimate becomes once every $\gtrsim$ 12 years, in very good agreement with the data from PSR J1023+0038.  As the \fermi\ mission continues it is thus likely that other systems may transition, and monitoring the flux is important not only for known redback systems but also of unassociated sources with pulsar-like spectra from which pulsations have not yet been detected.  Previous studies have defined pulsar-like sources to be non-variable with significantly curved spectra; however, given the measured spectral variability of \mypsr\ during the LMXB phase it seems wise to relax the first criterion when searching for similar systems.

\acknowledgments

\begin{center}
\emph{ACKNOWLEDGMENTS}
\end{center}

The \fermi\ LAT Collaboration acknowledges generous ongoing support from a number of agencies and institutes that have supported both the development and the operation of the LAT as well as scientific data analysis.  These include the National Aeronautics and Space Administration and the Department of Energy in the United States, the Commissariat \`a l'Energie Atomique and the Centre National de la Recherche Scientifique / Institut National de Physique Nucl\'eaire et de Physique des Particules in France, the Agenzia Spaziale Italiana and the Istituto Nazionale di Fisica Nucleare in Italy, the Ministry of Education, Culture, Sports, Science and Technology (MEXT), High Energy Accelerator Research Organization (KEK) and Japan Aerospace Exploration Agency (JAXA) in Japan, and the K.~A.~Wallenberg Foundation, the Swedish Research Council and the Swedish National Space Board in Sweden.

Additional support for science analysis during the operations phase is gratefully acknowledged from the Istituto Nazionale di Astrofisica in Italy and the Centre National d'\'Etudes Spatiales in France.

The Parkes radio telescope is part of the Australia Telescope which is funded by the Commonwealth Government for operation as a National Facility managed by CSIRO.

Portions of this research performed at the Naval Research Laboratory are sponsored by NASA DPR S-15633-Y.

\textit{Facilities:} \facility{Fermi(LAT)}, \facility{Parkes}

\bibliographystyle{apj}
\bibliography{references_new}

\end{document}